\documentclass[aps,epsf]{iopart}
\usepackage{graphics}
\usepackage{graphicx}
\usepackage{epsfig}

\def\beq{\begin{equation}}
\def\eeq{\end{equation}}
\def\bea{\begin{eqnarray}}
\def\eea{\end{eqnarray}}

\begin{document}
\input epsf.tex

\title{Searching for Massive Black Hole Binaries in the first Mock LISA Data Challenge}
\author{Neil J. Cornish$^{1}$ and Edward K. Porter$^{1,2}$} 
\address{1. Department of Physics, Montana State University, Bozeman, 59717, MT, USA.\\
2. Max Planck Institut f\"{u}r Gravitationsphysik, Albert Einstein Institut, Am M\"{u}hlenberg 1, Golm, Germany}
\vspace{1cm}
\begin{abstract}
\noindent The Mock LISA Data Challenge is a worldwide effort to solve the LISA data analysis problem.
We present here our results for the Massive Black Hole Binary (BBH) section of Round 1. Our
results cover Challenge 1.2.1, where the coalescence of the binary is seen, and Challenge 1.2.2, where
the coalescence occurs after the simulated observational period. The data stream is composed of Gaussian
instrumental noise plus an unknown BBH waveform.  Our search algorithm is based on a variant of the
Markov Chain Monte Carlo method that uses Metropolis-Hastings sampling and thermostated
frequency annealing. We present results from the training data sets and the blind data sets.
We demonstrate that our algorithm is able to rapidly locate the sources, accurately recover the source
parameters, and provide error estimates for the recovered parameters.
\end{abstract}

\maketitle

\section{Introduction}
Massive black hole binaries (BBH) are expected to be one of the strongest candidate sources for the
Laser Interferometer Space Antenna, LISA, a joint ESA-NASA mission that will search for gravitational
waves (GW) is the frequency bandwidth $10^{-5}\leq f/{\rm Hz} \leq 1$~\cite{LISA}.  The detection of BBHs by
LISA is important for many reasons.  Firstly, it will allow us to carry out a test of gravity in the highly
nonlinear strong-field regime~\cite{Ryan, colhugh, bercord}.  Secondly, it will allow us, in conjunction
with other astronomical methods, to investigate such things as galaxy interactions and mergers out to very
high redshift $(z\geq 10)$.  It will also allow us to test galaxy formation models such as Hierarchical formation,
where it is believed that modern day galaxies were formed from the merger of smaller ``seed'' galaxies.  Due to
their high masses, the inspiral phases for these systems occur at frequencies which are unavailable to the ground
based detectors due to their low frequency cutoffs.  Also, unlike galactic binaries and Extreme Mass Ratio
Inspirals (EMRIs), the BBHs are very clean sources with detectable signal to noise ratios (SNR) of order
$\sim 10 - 1000$s.  This means that we will not have to deal with confusion noise between
sources~\cite{portercornish06c}, something which is very important in the search for galactic binaries and
to a lesser extent in the search for EMRIs.

Here we describe our analysis of the BBH component of the Mock LISA Data Challenge~\cite{MLDC}. Our search algorithm is
a variant of the Markov Chain Monte Carlo method (MCMC) which has been described in
References~\cite{portercornish06c,portercornish06a,portercornish06b}.  The algorithm uses a mixture of frequency annealing
with thermostated heat, simulated annealing, plus a 5-D exploration of the posterior distributions for the
search parameters  (we use a generalized ${\cal F}$-statistic to automatically search over the distance,
inclination, polarization and initial phase).
We refer the reader to Reference~\cite{portercornish06c} for a full description of the algorithm.

In our earlier work we used the Low Frequency Approximation (LFA)~\cite{cutler98} to model the instrument response, but
to our surprise this proved to be insufficiently accurate for the MLDC data sets where the full detector response
is used. Since the maximum frequencies of the injected signals were below 2 mHz, we had expected the LFA to be
adequate, but when running on the training data sets we found systematic offsets in many of the parameters. The
parameter recovery improved significantly when we upgraded our instrument response to the Rigid
Adiabatic Approximation (RAA)~\cite{crp}, which includes finite armlength effects. Our interpretation of this finding
is that while the differences between the LFA and RAA are small at 2 mHz, the differences are amplified by
the very large contribution to the signal to noise ratio that comes from from the final cycles of the inspiral.
Indeed, we suspect that the remaining small systematic offsets in the recovered masses can be traced to
the approximate nature of the RAA.

We modified our barycenter waveforms and signal tapers to agree with those used to inject the simulated
signals~\cite{ch1}, but rather than searching over initial phase (the phase parameter used to generate
the waveforms), we continued our earlier practice of searching over the phase at coalescence, as we
have found this to give better acceptance rates in the search chains.

The organization of the rest of the paper is as follows : In Section~\ref{sec:method} we present a very short 
discussion of the search algorithm used.  We define the Metropolis-Hastings sampling, the frequency annealing
with thermostated heat and the simulated annealing scheme used. Section~\ref{sec:challenge} contains a presentation
of results for both the training and blind data sets for the challenges.


\section{The Search Algorithm.}\label{sec:method}
Our search algorithm, which is again explained in more detail in \cite{portercornish06c} is based on a Metropolis-Hastings
sampling, which is the central engine of the standard Markov Chain Monte Carlo method (MCMC).  Our method incorporates
both simulated and frequency annealing schemes at various stages in the search.  The algorithm uses a number of proposal
distributions to jump around the parameter space, as well as a maximization over the time to coalescence.  
To quickly summarize : Starting with the signal $s(t)$ and some initial template $h(t,\vec{x})$ the (un-normalized)
posterior density at $\vec{x}$ is computed. We then draw from a proposal distribution and propose a jump to another
point in the space $\vec{y}$.  The posterior and proposal densities at $\vec{x}$ and $\vec{y}$ are then compared
by forming the the Metropolis-Hastings ratio
\begin{equation}
H = \frac{\pi(\vec{y})p(s|\vec{y})q(\vec{x}|\vec{y})}{\pi(\vec{x})p(s|\vec{x})q(\vec{y}|\vec{x})}.
\end{equation}
Here $\pi(\vec{x})$ are the priors of the parameters, $p(s|\vec{x})$ is the likelihood, and $q(\vec{x}|\vec{y})$ is
the proposal distribution.  This jump is then accepted with probability $\alpha = min(1,H)$, otherwise the chain
stays at the proposal point. 

The likelihood was calculated using a generalized ${\cal F}$-Statistic~\cite{jks}, which allows for 
an analytic extremization over the extrinsic parameters: luminosity distance; orbital inclination, polarization
and orbital phase at coalescence. Consequently, the Markov Chain portion of our search returns a marginalized
posterior distribution. For future Challenges, we intend to follow the initial detection with a full parameter
search so as to obtain the full posterior.

In the first section of the search we use a frequency annealing scheme to speed up the search.
The search templates and the inner products used to compute the ${\cal F}$-Statistic are
terminated at a cut-off frequency $f_{\rm cut}$, which is initially set at $4\times10^{-5}$ Hz.
The cut-off frequency is then increased as the chain progresses according to
\begin{equation}
f_{\rm cut}= \left\{ \begin{array}{ll} 10^{-B(1-i)}f_{\rm max} & f< f_{\rm max} \\ \\ 
f_{\rm max} & f \geq f_{\rm max}  \end{array}\right.,
\end{equation}
where $i$ is the number of steps in the chain, $B$ is a growth parameter and $f_{\rm max}$ is the maximum
frequency the signals could reach give the priors on the masses {\it etc.}. As the cut-off frequency
is incremented, more of the BBH signal is revealed, and the SNR of the best fit templates increases.
Thus, frequency annealing acts in a similar way to traditional simulated annealing, but with the
added benefit of saving in the cost of the template generation and ${\cal F}$-Statistic computation.
In testing we found that the search chains would sometimes lock onto secondary maxima during the
frequency annealing phase, so we introduced a ``thermostating'' procedure to control the effective
SNR of the recovered signals. This was done by multiplying the noise spectral density in the
noise weighted inner products by a ``heat'' factor $\beta$, which was adjusted based on the
SNR of the current template:
\begin{equation}
\beta = \left\{ \begin{array}{ll} 1.0 & 0\leq {\rm SNR}\leq 20 \\ \\ \left(\frac{{\rm SNR}}{20}\right)^{2} & {\rm SNR} > 20  \end{array}\right. ,
\end{equation}
Once the frequency annealing stage was completed, the chain was cooled using the simulated annealing scheme 
\begin{equation}
\beta = \left\{ \begin{array}{ll} 10^{-\xi\left(1-\frac{j}{N_{c}}\right)} & 0\leq j\leq N_{c} \\ \\ 1 & j > N_{c}  \end{array}\right.,
\end{equation}
with $\xi=\log_{10}\beta_{\rm max}$ where $\beta_{\rm max}$ is the heat factor at the end of the frequency annealing
stage. The index $j$ counts from the end of the frequency annealing, and the cool down lasts $N_c$ steps. 
A standard MCMC exploration of the marginalized posterior distribution function (PDF) commences once $\beta = 1$
and continues for $N_e$ steps. Finally, over the course of $N_f$ steps, we freeze the chain to a heat
of $\beta = 0.01$ to aid in the extraction of Maximum Likelihood Estimates (MLEs) for each of the parameters.


\section{Conducting the Challenge.}\label{sec:challenge}
The MLDC for binary black holes was broken into two parts. Challenge 1.2.1 had a source that coalesced inside the observation period,
while Challenge 1.2.2 had a source coalescing outside of the observation period.  The data sets provided consisted of
$2^{21}$ data points  sampled every 15 seconds giving approximately one year of data.

\subsection{Challenge 1.2.1.}
Priors on the source parameters were given for each challenge. For Challenge 1.2.1 we were told that
mass ranges were restricted such that $m_{1} \in [1,5]\times 10^{6} M_{\odot}$ and $m_{2} = m_{1} / x$,
where $1\leq x\leq 4$, and that the time to coalescence lay in the range $t_c \in [5,7]$ months. We were also
told that the system would have a signal to noise in the range $450\leq {\rm SNR} \leq 500$ in one interferometer.
No priors were given on the other source parameters. In addition to the blind data set, a training data set
was made available with parameters drawn from the same set of priors.

\begin{figure}[t]
\begin{center}
\epsfig{file=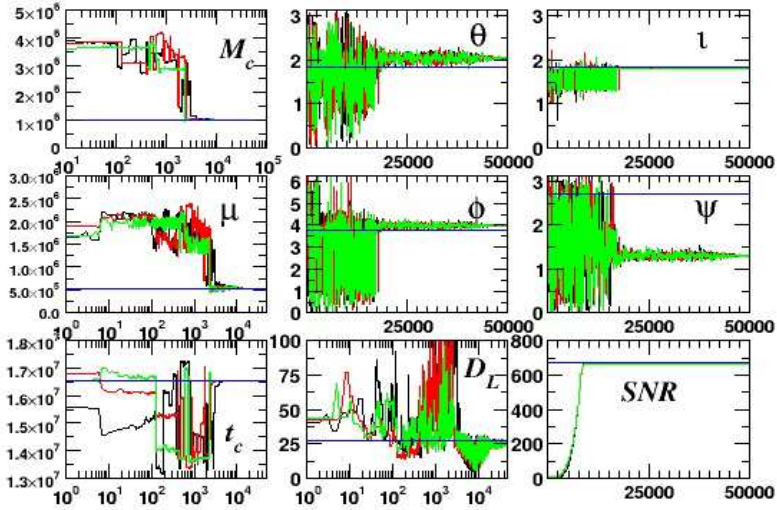, width=4.5in, height=3.5in}
\end{center}
\vspace{-8mm}
\caption{A plot of three search chains for eight of the nine parameters in the Challenge 1.2.1 training set.
The solid lines in each cell denote the true value.}
\label{fig:searcht121}
\end{figure}

For the 1.2.1 training data set we ran several 50,000 point chains.  These chains were composed of a 10,000 point frequency
annealing search, a 10,000 point simulated annealing phase cooldown, a 20,000 point MCMC chain to explore the PDFs,
and a final 10,000 point freezing of the chain to extract the MLEs. In Figure~\ref{fig:searcht121} we plot three
different search chains for eight of the nine waveform parameters in the 1.2.1 training set.  We have omitted
the search chain for the initial phase. We plot the extrinsic parameter chains even though their values were
determined by analytical extremization at each point in the chain. While the angular variables employ a linear
scale for the number of steps in the search, the other parameters are plotted against a logarithmic scaling
to highlight the early convergence that occurs for the parameters such as the time to coalescence and the masses.
The training data set provided for Challenge 1.2.1 proved to be more challenging than similar test cases
we generated for ourselves. In particular, the sky position of the source was not recovered to the accuracy
predicted by a Fisher matrix estimate of the measurement errors. We attribute this to an unfortunate alignment of
the source with respect to the LISA at the time of coalescence. At coalescence the motion of the
detector turns out to be perpendicular to line of sight to the source, so there is little or no Doppler
shift of the gravitational waves right at the time when most of the SNR is accumulating. Thus, directional
information gets less weighting than is typical, and the small differences between the RAA and the full
response used to generate the data sets is amplified. The problem went away when we
tested our search algorithm on the same source using data that was simulated with the RAA.
In future Challenges we plan to use a full detector response model to
generate the search templates.

\begin{figure}[t]
\begin{center}
\epsfig{file=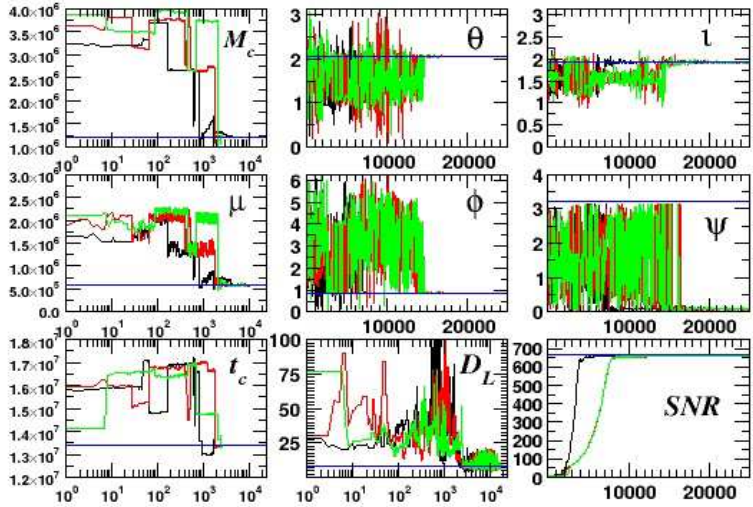, width=4.5in, height=3.5in}
\end{center}
\vspace{-8mm}
\caption{A plot of three search chains for eight of the nine parameters for the blind Challenge 1.2.1 data set.  The
solid lines in each cell denote the true value.}
\label{fig:searchb121}
\end{figure}

The sky location for the 1.2.1 blind data set proved to be easier to pin down and we were able to get away with
shorter search chains of 25,000 points. These chains were composed of a 5,000 point frequency annealing search,
a 5,000 point simulated annealing phase cooldown, a 10,000 point MCMC chain to explore the PDFs at unit heat,
ending with a 5,000 point freezing of the chain to extract the MLEs.  We have plotted the search chains for the
blind data set in Figure~\ref{fig:searchb121}, along with the values of the injected source parameters which
were revealed after we had submitted our results. We see that the chains typically find the three most important
parameters $(M_{c},\mu,t_{c})$ in around 1000 steps. Once again the sky positions took longer to converge, but
in this instance the chains converged on the injected parameter values. The search also accurately recovered
the extrinsic parameters, save for the initial phase. The failure to recover the initial phase was due to
a bug in our ${\cal F}$-Statistic routine that we overlooked when working on the training data. We should also
mention that while the recovered value for the polarization $\psi$ is off by $\pi$ from what was injected,
the polarization angle is only defined up to multiples of $\pi$ so the solution is physically identical.

\begin{figure}[t]
\begin{center}
\epsfig{file=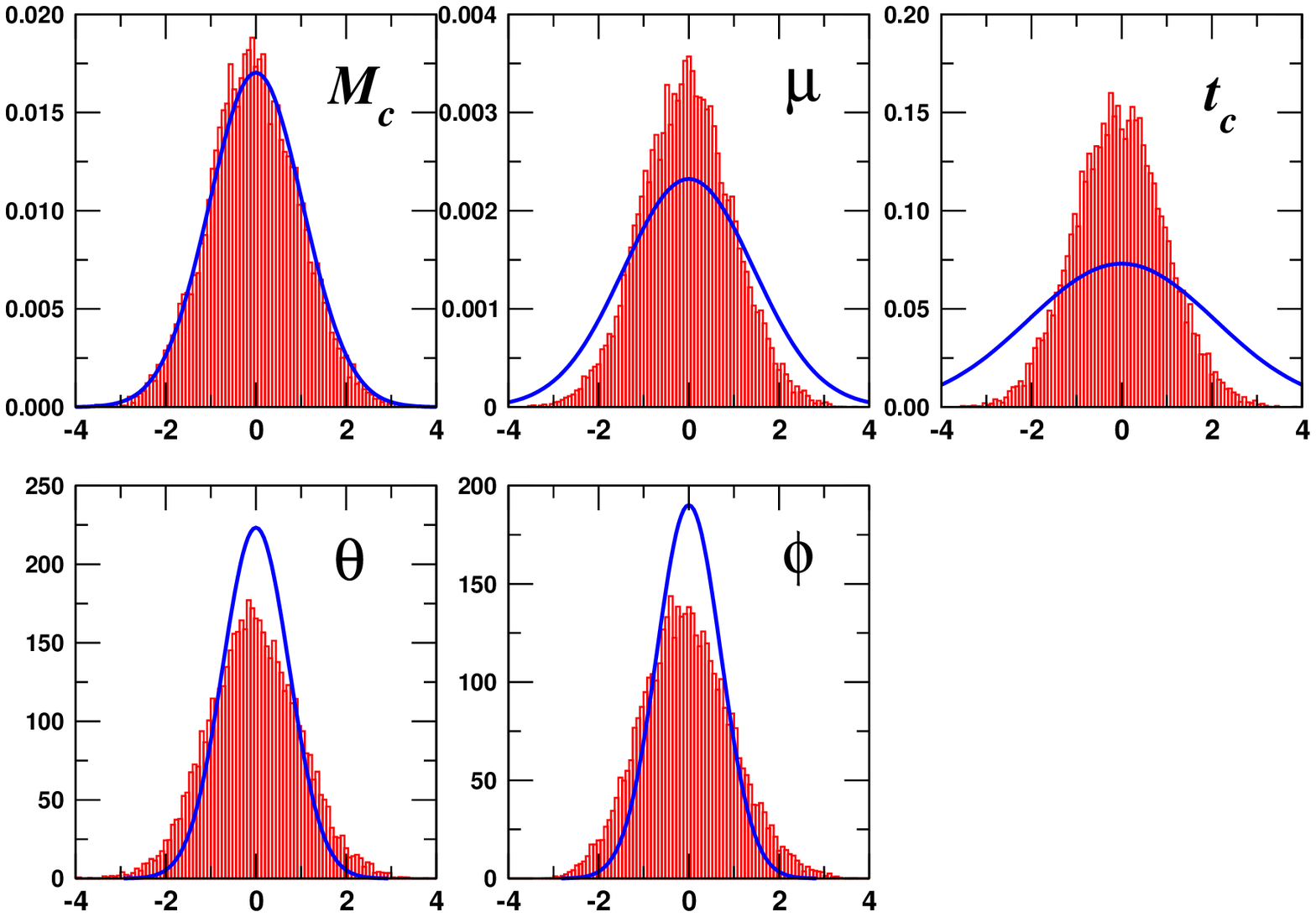, width=4.5in, height=3.5in}
\end{center}
\caption{A comparison of the marginalized PDFs from the MCMC chains for the intrinsic parameters of 1.2.1 against
the Fisher prediction (solid line) from the mean of the chain.  The means of the chain have been subtracted and the
values scaled by the square root of the variances of the chains.}
\label{fig:histb121}
\end{figure}

In Figure~\ref{fig:histb121} we have plotted the marginalized PDFs for the extrinsic parameters based on the
merged MCMC chains from each different run.  The solid line in each cell is the Fisher matrix prediction for
the marginalized posteriors. While the chains we not long enough to fully characterize the posteriors, we see that
the merged chains show good agreement with Fisher predictions for $(M_{c},\theta,\phi)$. The Fisher
predictions for $(\mu,t_{c})$ do not agree very well with the MCMC results. At this time we do not have an explanation for the disagreement.

In Table~\ref{tab:ch121} we compare that values of the key file against the MLEs for the training set (top)
and the blind set (bottom).  We also quote the 1-$\sigma$ error estimation from the Fisher matrix, and the error
in multiples of the 1-$\sigma$ error estimates from the Fisher matrix.  In the table, $t_{c}$
is recorded in seconds.  While again we did not search for them, the MLEs obtained for $(\iota, \ln D_{L}, \psi)$
give errors of $(3.86\times10^{-3}, 6.62\times10^{-2}, 1.43)$ and $(1.7\times10^{-2}, 4.15\times10^{-2}, -2.7\times10^{-4})$ for the training and blind data setsrespectively. 
The $\psi$ value is obtained from converting $\psi\rightarrow\psi+\pi$.  For the blind data set, the source had a
combined SNR of 664.78, while we recovered a SNR of 658.38 .  Again the mismatch is due, we feel, to a phasing issue
in the code which has since been rectified.  Each run took approximately 24 hours on a single Mac G5 processor.

\begin{table}
\begin{center}
\begin{tabular}{|c|c|c|c|c|}\hline\hline
&  & & &  \\ 
& $\lambda^{X}$ & $\lambda^{MLE}$ & $\sigma^{Fisher}_{\lambda^{X}}$ & $n\sigma^{Fisher}_{\lambda^{X}}$ \\
&  & & & \\ \hline
&  & & & \\ 
$M_{c}$ & $\,\,\,\,1.023866\times10^{6}\,\,\,\,$ & $\,\,\,\,1.024227\times10^{6}\,\,\,\,$ & 45.06 & -8.01 \\ 
$\mu$ & $5.373042\times10^{5}$ & $5.38975\times10^{5}$ & 269.55 & -6.19 \\
$\theta$ & 1.8339 & 2.0488 & $6.673\times10^{-2}$& -3.22 \\
$\phi$  & 3.7945 & 0.3.9981 & $4.414\times10^{-2}$& -4.61 \\
$t_{c}$ & $1.6545493\times10^{7}$ & $1.6545493\times10^{7}$ & $13.67$ & 5.43 \\
&  & & &   \\ \hline
&  & & &   \\ 
$M_{c}$ & $\,\,\,\,1.20859\times10^{6}\,\,\,\,$ & $\,\,\,\,1.2087\times10^{6}\,\,\,\,$ & $23.78$& -5.004 \\ 
$\mu$ & $5.81196\times10^{5}$ & $5.818\times10^{5}$ & 173.35 & -3.5 \\
$\theta$ & 2.0631 & 2.0619 & $1.81\times10^{-3}$& 0.63 \\
$\phi$  & 0.8658 & 0.8645 & $2.12\times10^{-3}$& 0.63 \\
$t_{c}$ & $1.3374027\times10^{7}$ & $1.3374031\times10^{7}$ & $5.53$ & -0.62 \\
&  & & & \\ 
\hline\hline
\end{tabular}
\end{center}
\caption{This table compares the injected parameter values, the MLE for each parameter, the $1\sigma$ error predicted
by the Fisher matrix at the injected values, and the difference between the MLEs and the injected values in multiples
of the Fisher $1\sigma$ error estimate the Challenge 1.2.1 training set (top) and blind set (bottom).}
\label{tab:ch121}
\end{table}

\subsection{Challenge 1.2.2.}
The priors for Challenge 1.2.2 gave a time to coalescence of $400\pm 40$ days and masses chosen such
that $m_{1} \in [1,5]\times 10^{6} M_{\odot}$ and again $m_{2} = m_{1} / x$, where $1\leq x\leq 4$.
No priors were given on the other six parameters, save that the source would have an SNR in the range of
$20\leq SNR \leq 100$ in one interferometer. For both the training and blind sets, we used 25,000 point
search chains.  These chains were composed of a 5,000 point frequency annealing search, a 5,000 point simulated
annealing phase cooldown, a 10,000 point MCMC chain to explore the PDFs at unit heat, ending with a
5,000 point freezing of the chain to extract the MLEs.

\begin{figure}[t]
\begin{center}
\epsfig{file=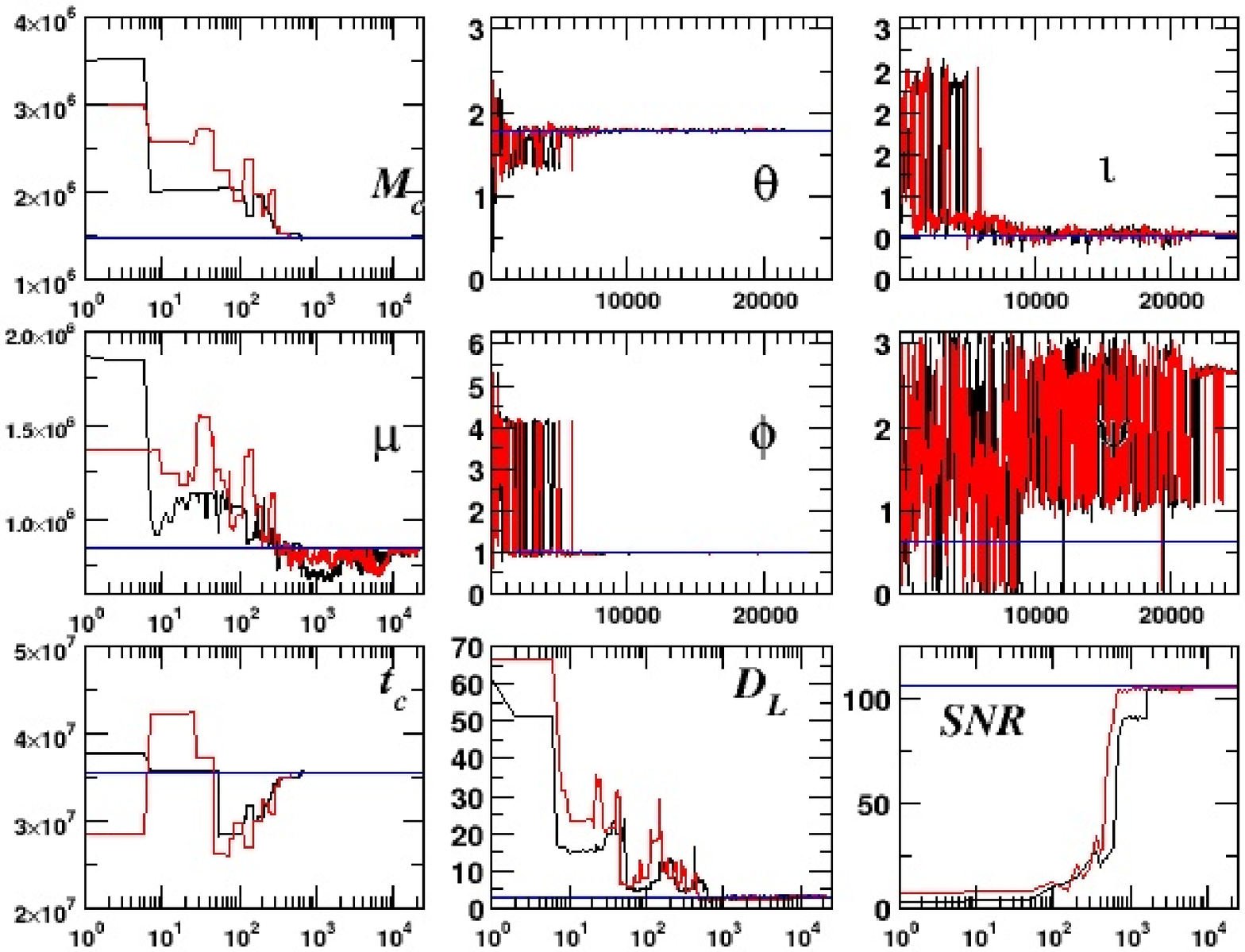, width=4.5in, height=3.5in}
\end{center}
\vspace{-8mm}
\caption{A plot of three search chains for eight of the nine parameters of the Challenge 1.2.2 training data set.
The solid lines in each cell denote the true value.}
\label{fig:searcht122}
\end{figure}

In Figure~\ref{fig:searcht122} we plot three search chains for the training data set.  The chirp mass and time to
coalescence are recovered in under 1000 steps. For this particular source, the chains also rapidly converge on
the correct sky position.  Again, while not explicitly searched for, the search correctly recovered the
inclination of the orbit and luminosity distance.  However, the recovered polarization angle is not very good.
We should note here that we have mapped the key file value back into a 0 to $\pi$ range.

\begin{figure}[t]
\begin{center}
\epsfig{file=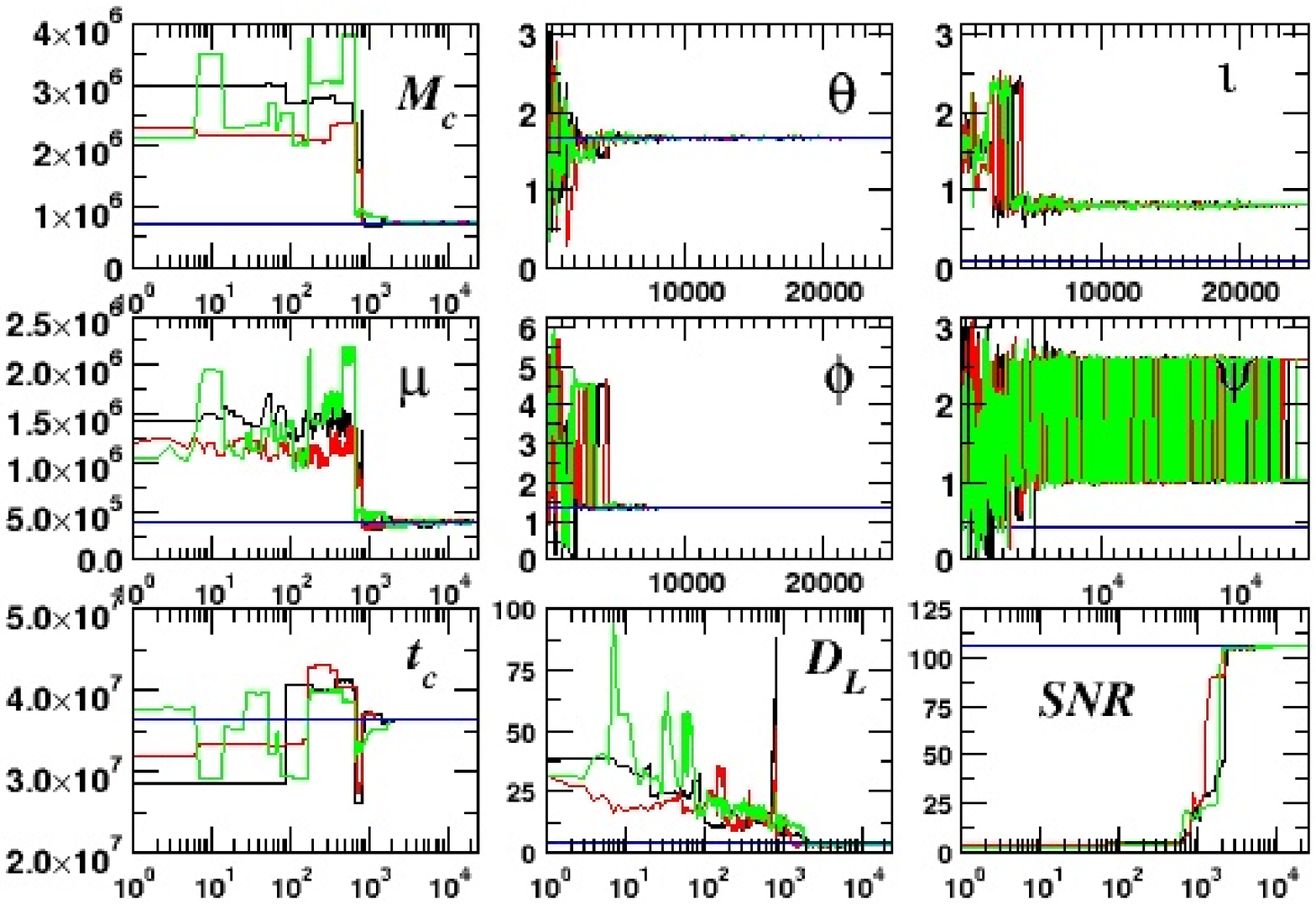, width=4.5in, height=3.5in}
\end{center}
\vspace{-8mm}
\caption{A plot of three search chains for eight of the nine blind parameters in 
Challenge 1.2.2.  The solid lines in each cell denote the true value.}
\label{fig:searchb122}
\end{figure}

\begin{figure}[t]
\begin{center}
\epsfig{file=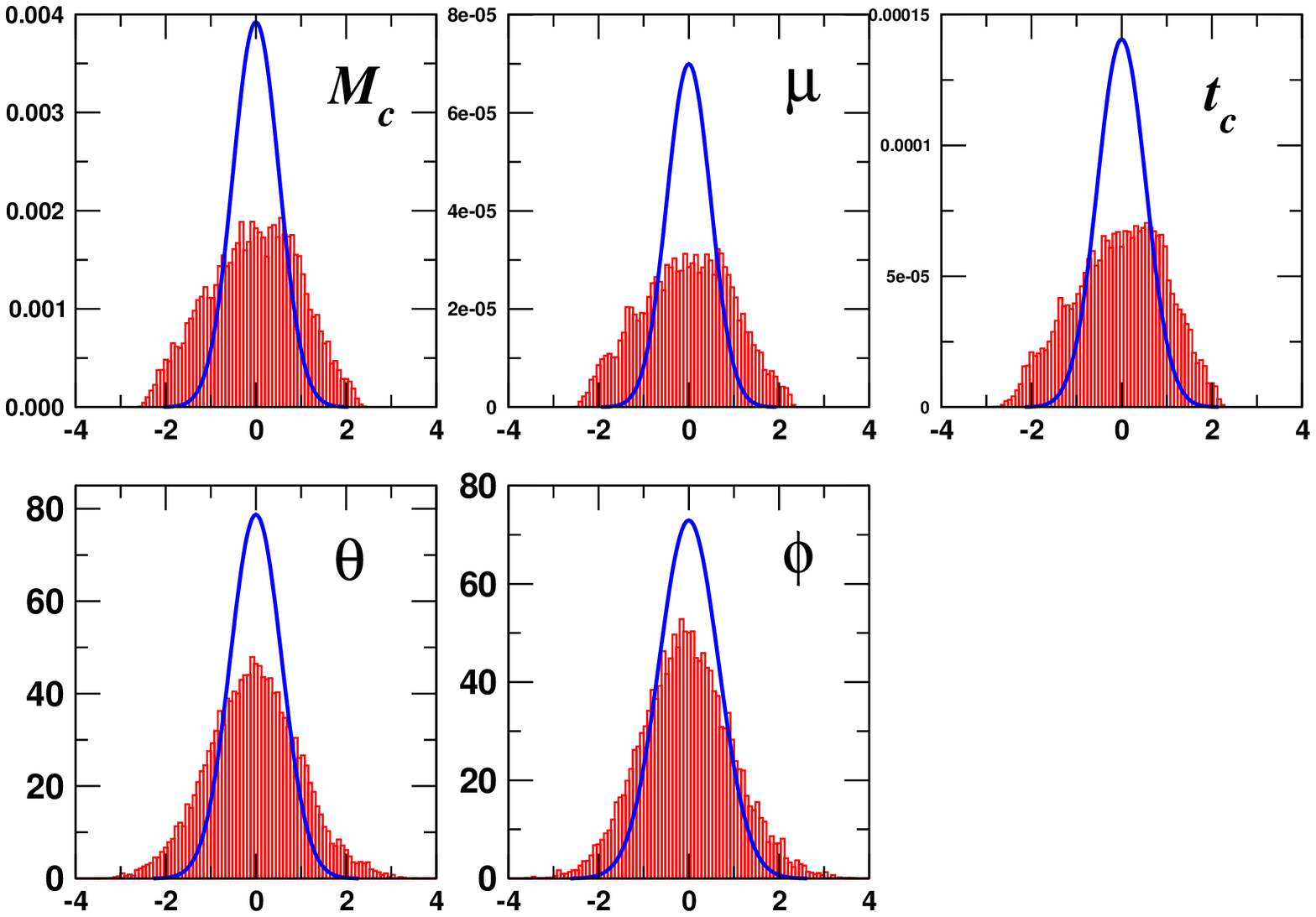, width=4.5in, height=3.5in}
\end{center}
\caption{A comparison of the marginalized PDFs from the MCMC chains for the intrinsic
parameters of 1.2.2 against the Fisher prediction (solid line) from the mean of the chain.
The means of the chain have been subtracted and the values scaled by the square root of the variances of the chains.}
\label{fig:histb122}
\end{figure}

In Figure~\ref{fig:searchb122} we plot three search chains for the blind data set.
All five intrinsic parameters lock-in after a few thousand steps.  Except for the luminosity distance, the extrinsic
parameters are far from the true values.  We attribute this to the fact the phase at coalescence is essentially
undetermined for systems where we do not see coalescence. In Figure~\ref{fig:histb122} we plot
the marginalized PDFs for the extrinsic parameters based on the merged MCMC chains from multiple runs.
The solid line in each cell is the prediction of the Fisher matrix at the mean of the chain.  There is a
discrepancy between the Fisher prediction and the posteriors from the chains. A visual inspection of
the chains indicates a large autocorrelation that extends over thousands of points. This slow mixing
of the chains implies that we would need to run much longer MCMC segments in order to get meaningful
posterior distributions to compare to the Fisher predictions.

In Table~\ref{tab:ch122} we compare that values of the injected parameters against the MLEs.  We again quote
the 1-$\sigma$ error estimation from the Fisher matrix and the error in multiples of the Fisher matrix
based on the MLEs.  Also, based on the ${\cal F}$-Statistic search for the intrinsic parameters, the
MLEs obtained for $(\iota, \ln D_{L}, \psi)$ give errors of $(-0.0313, 0.0124, 1.09)$ and $(-0.71, 0.273, -2.16)$ for the
training and blind data sets respectively.  The source had a combined SNR of 106.54, while we recovered
a SNR of 105.72 .  Each run took approximately 6 hours on a single 2 GHz Mac G5 processor.

\begin{table}
\begin{center}
\begin{tabular}{|c|c|c|c|c|}\hline\hline
&  & & &   \\ 
& $\lambda^{X}$ & $\lambda^{MLE}$ & $\sigma^{Fisher}_{\lambda^{X}}$ & $n\times\sigma^{Fisher}_{\lambda^{X}}$\\ 
&  & & &   \\ \hline
&  & & &   \\ 
$M_{c}$ & $\,\,\,\,1.47637\times10^{6}\,\,\,\,$ & $\,\,\,\,1.4765\times10^{6}\,\,\,\,$ & 209.5 & -0.66 \\ 
$\mu$ & $8.4183\times10^{5}$ & $8.4748\times10^{5}$ & 8466.8 & -0.67 \\
$\theta$ & 1.7802 & 1.7862 & $4.09\times10^{-3}$& 1.47 \\
$\phi$  & 0.9737 & 0.9904 & $4.22\times10^{-3}$& -3.95 \\
$t_{c}$ & $3.559836\times10^{7}$ & $3.559978\times10^{7}$ & 2361.8 & -0.6 \\
&  & & &   \\ \hline
&  & & & \\ 
$M_{c}$ & $\,\,\,\,7.4146\times10^{5}\,\,\,\,$ & $\,\,\,\,7.4169\times10^{5}\,\,\,\,$ & 97.78 & -2.35 \\ 
$\mu$ & $3.848\times10^{5}$ & $3.989\times10^{5}$ & 5,273 & -2.66\\
$\theta$ & 1.6947 & 1.6758 & $5.1\times10^{-3}$& 3.73 \\
$\phi$  & 1.3674 & 1.362 & $5.27\times10^{-3}$& 1.03 \\
$t_{c}$ & $3.63076\times10^{7}$ & $3.63142\times10^{7}$ & 2,827 & -2.33 \\
&  & & &   \\ 
\hline\hline
\end{tabular}
\end{center}
\caption{This table shows the injected parameter values, the MLE for each parameter, the $1\sigma$ error
predicted by the Fisher matrix at the injected values and the difference between the MLEs and the
injected values in multiples of the Fisher $1\sigma$ error estimate for the Challenge 1.2.2 training set (top)
and blind data set (bottom).}
\label{tab:ch122}
\end{table}

\section{Conclusion.}
The Mock LISA Data Analysis Challenge data sets for binary black hole inspirals proved to be a valuable testing
ground for our Metropolis-Hastings search algorithm. Overall the algorithm performed very
well, recovering the injected parameters to an accuracy largely consistent with the theoretical error margins.
There were also some surprises, such as needing to go over to a more accurate treatment of the instrument
response in our template generation, and a problem with our conventions when extracting the initial
orbital phase. Work is now underway on the much more complicated multi-source Challenge 2 data sets,
and early results from runs on the training data look very promising.

\section*{Acknowledgments}
This work was supported at MSU by NASA Grant NNG05GI69G.

\pagebreak
\appendix


\section*{References}
\begin{thebibliography}{99}
\bibitem{LISA} 
Bender~P et al., \emph{LISA pre-phase A report} (1998)

\bibitem{Ryan} Ryan~F~D, Phys. Rev. D{\bf 56}, 1845 (1997).

\bibitem{colhugh} Collins~N~A \& Hughes~S~A, Phys. Rev. D{\bf 69}, 124022 (2004).

\bibitem{bercord} Berti~E, Cardoso~V \& Will~C~M, Phys. Rev. D{\bf 73}, 064030 (2006).

\bibitem{portercornish06c} Cornish~N.J. \& Porter~E.K., gr-qc/0612091

\bibitem{MLDC} Arnaud~K~A, Babak~S, Baker~J~G, Benacquista~M~J, Cornish~N~J, Cutler~C, Larson~S~L, Sathyaprakash~B~S, Vallisneri~M,  Vecchio~A, Vinet~J-Y, gr-qc/0609105 (2006)

\bibitem{mldc} http://astrogravs.nasa.gov

\bibitem{portercornish06a} Cornish~N.J. \& Porter~E.K., Class. Quant. Grav. {\bf 23} S761 (2006)

\bibitem{portercornish06b} Cornish~N.J. \& Porter~E.K., Phys. Rev. D{\bf 75}, 021301 (2007).

\bibitem{cutler98} Cutler~C, Phys. Rev. D {\bf 57}, 7089 (1998)

\bibitem{crp} Rubbo~L~J, Cornish~N~J \& Poujade~O, Phys. Rev. D{\bf 69} 082003 (2004)

\bibitem{ch1} MLDC Challenge 1 Omnibus Document, {\rm http://svn.sourceforge.net/viewvc/lisatools/Docs/challenge1.pdf}.

\bibitem{jks} Jaranowski~P, Kr\'{o}lak~A and Schutz~B~F, Phys.~Rev.~D {\bf 58}, 063001 (1998)
\end{thebibliography}
\end{document}